\documentclass{article}
\usepackage{amssymb}
\usepackage{amsfonts}
\usepackage[a4paper]{geometry}
\usepackage{amsmath}

\input{tcilatex}
\begin{document}

\title{ Geometrical Amplitude factors in the the adiabatic evolution}
\author{Mustapha Maamache \\
Laboratoire de Physique Quantique et Syst\`{e}mes Dynamiques,\\
Facult\'{e} des Sciences, Universit\'{e} Ferhat Abbas S\'{e}tif 1, S\'{e}tif
19000, Algeria}
\maketitle

\begin{abstract}
In a quantum system initially in the $n$-th eigenstate, an adiabatic
evolution of the Hamiltonian ensures that the system remains in the
corresponding instantaneous eigenstate while acquiring a phase factor. This
phase has two components: one resulting from standard time evolution and
another associated with the dependence of the eigenstate on the varying
Hamiltonian, known as the Berry phase. In this work, we explore the concept
of geometric amplitudes in the context of a Hermitian Hamiltonian with
imaginary eigenvalues. We introduce the notion of geometric amplitude and
provide a novel derivation of this concept. Our study reveals that a system
undergoing cyclic evolution under adiabatic conditions acquires an
additional amplitude factor of purely geometric origin. To illustrate this
idea, we apply it to a concrete case: a generalized harmonic oscillator.


In memory of my mother Djabou Zoulikha and my uncle Boukhari Miloud.
\end{abstract}

\section{Introduction}

Quantum mechanics (QM), the fundamental theory describing microscopic
phenomena, was developed with rigorous mathematical formulations, ensuring
its applicability across nearly all branches of physics. One of the most
crucial aspects of QM is predicting the time evolution of a Hamiltonian
system, which is essential for addressing many practical physical problems.

Despite its remarkable success and widespread applications in modern
science, QM also has certain limitations. For instance, a fully consistent
formulation of QM is typically confined to self-adjoint (in the Dirac
sense), i.e., Hermitian quantum systems \cite{1}. This constraint ensures
that energy and other observable quantities take real values, with a
complete set of orthonormal eigenfunctions, and that the time evolution
remains unitary. Given these fundamental principles, any study of QM that
does not account for Hermitian operators remains significantly limited.

In the year 1998, Carl M Bender and his collaborators have found a certain
class of Non Hermitian Hamiltonians which holds an entire real discrete
spectrum \cite{Ben98, Ben02}. The reality of the spectrum is shown to be a
direct consequence of unbroken symmetry under \ combined parity ($P$) and
time reversal ($T$) transformations . Since then such non-Hermiuian $PT$%
-symmetric systems acquire a great importance in quantum theory and have
been studied rigorously. Historically, non-Hermitian systems attracted
attention in the 1990s, when Carl Bender and his colleagues challenged the
well-known axiom of QM saying that Hermicity is not necessary to observe
real eigenvalue spectra \cite{Ben98, Ben02}. This discovery encouraged many
researchers to revisit the Hermitian theory of QM. Mostafazadeh pointed out
that pseudo-Hermicity is the responsible for the real spectra of $PT$%
-symmetric Hamiltonians, which belong to a specific class of the general
family of pseudo-Hermitian Hamiltonians \cite{Mos02}.

As mentioned, a fundamental fact in quantum mechanics is that Hermitian (or
self-adjoint) operators have only real eigenvalues. However, one may
encounter Hermitian Hamiltonians with imaginary eigenvalues. A simple
example may illustrate the issue: Consider quantum mechanics on the real
line and the quantum mechanical operator $O=q^{3}p+pq^{3}\ $. This operator
is hermtian (or symmetric) but it still has an eigenvalue $\left( 
{\normalsize -i\lambda }\right) $\ with a square-integrable eigenfunction\ $%
\chi _{\lambda }(q)=1/|q|^{{\normalsize 3/2}}$e$^{{\normalsize -\lambda \ /4q%
}^{2}}$. Such cases involving imaginary eigenvalues have not been widely
explored in the literature and will be one of the key topics discussed in
this paper.

However, if the Hamiltonian $H$ \ satisfies $H=H^{\dagger }$\ but has
imaginary eigenvalues, then the time evolution operator will also have
imaginary eigenvalues. As a result, $exp(-iHt)$ will no longer a pure phase,
leading to a time-dependent evolution of the norm of quantum states.
Considering the eigenvalue equation:

\begin{equation}
H\left\vert n\right\rangle =iE_{n}\left\vert n\right\rangle ,\text{ \ }E_{n}%
\text{ real\ \ }
\end{equation}%
the solution of Schr\"{o}dinger equation, 
\begin{equation}
i\frac{\partial }{\partial t}\left\vert \Psi _{n}(t)\right\rangle =H\left(
t\right) \left\vert \Psi _{n}(t)\right\rangle ,  \label{Sch}
\end{equation}%
is given as 
\begin{equation}
\left\vert \Psi _{n}(t)\right\rangle =e^{E_{n}t}\left\vert n\right\rangle ,
\label{0}
\end{equation}%
where $\hbar =1$ through the text. These solutions does not preserve
unitarity, i.e., $\langle \Psi _{n}(t)\shortmid \Psi _{n}(t)\rangle \neq 1.$%
To address this issue, we will define a new scalar product by introducing a
metric $\eta $ that ensures the norm remains unitary.

When the Hamiltonian is explicitly time dependent the situation becomes more
complicated. One way to address this challenge is to use approximation
methods, such as the adiabatic theorem, first introduced by Ehrenfest \cite%
{31}. This theorem applies to quantum systems governed by a Hamiltonian that
is explicitly time-dependent but varies slowly. The hermiticity of the
operators plays a crucial role in ensuring the validity of its proof \cite{2}%
-\cite{5}. In the following, we introduce Berry's adiabatic concept and the
notion of geometric phase.

\ In 1984 Berry published a paper \cite{6} which has until now deeply
influenced the physical community. Therein he considers cyclic evolutions of
systems under special conditions, namely adiabatic ones. He finds that an
additional phase factor occurs in contrast to the well known dynamical phase
factor. This phenomenon can be described by "global change without local
change". Berry points out the geometrical character of this phase which is
not be negligible because of its nonintegrable character. Since this much
work has been done to this issue and the so called Berry phase is now well
established, theoretically as well as experimentally.

The adiabatic theorem, a fundamental result in quantum theory, states that
for a Hamiltonian $H(t)$ that varies slowly over time, the solution of the
time-dependent Schr\"{o}dinger equation can be approximated using the
eigenstates $\left\vert n(t)\right\rangle $ of the instantaneous "snapshot"
Hamiltonian 
\begin{equation}
H(t)\left\vert n(t)\right\rangle =E_{n}(t)\left\vert n(t)\right\rangle ,
\end{equation}%
with energies $E_{n}(t)$ (\ \ $n=0,1,...$). This eigenvalue equation implies
no relation between the \ phases of the eigenstates $\left\vert
n(t)\right\rangle $ at different $t$.

The adiabatic theorem concerns states\ $\left\vert \Psi _{n}(t)\right\rangle 
$\ satisfying the time-dependent Schr\"{o}dinger equation and asserts that
if a quantum system with a time-dependent non degenerate Hamiltonian $%
H\left( t\right) $\ is initially in the $n$th eigenstates $\left\vert
n(0)\right\rangle $ of $\ H\left( 0\right) $, and if $\ H\left( t\right) $\
evolves slowly enough, then the state at time $t$\ remains in the $n$th\
instantaneous eigenstates of $H\left( t\right) $\ up to a multiplicative
phase factor.

$\bigskip $This phase factor%
\begin{equation}
\ \ \varphi _{n}\left( t\right) =-\int_{0}^{t}dt^{\prime }\left\langle
n(t^{\prime })\left\vert H(t^{\prime })-i\hbar \frac{\partial }{\partial
t^{\prime }}\right\vert n(t^{\prime })\right\rangle
\end{equation}%
consists of two contributions: a dynamical phase and a geometric phase.
Consequently, the quantum state $\left\vert \Psi _{n}(t)\right\rangle $ can
be expressed as:

\begin{equation}
\left\vert \Psi _{n}(t)\right\rangle =e^{i\int_{0}^{t}dt^{\prime
}\left\langle n(t^{\prime })\left\vert i\frac{\partial }{\partial t^{\prime }%
}-H(t^{\prime })\right\vert n(t^{\prime })\right\rangle }\left\vert
n(t)\right\rangle .
\end{equation}

\ The geometric phase \cite{6}- \cite{6a} appears in a wide range of
physical phenomena, from polarization changes in light propagating through
optical fibers \cite{14"}-\cite{15"} and the precession of a neutron in a
magnetic field \cite{16"}- \cite{24"} to the quantum dynamics of dark states
in atoms \cite{25"}. More recently, it has also played a role in topological
quantum computing \cite{26"}, as demonstrated with trapped ions \cite{27"}.
There have also been attempts to extend geometric phases to systems
described by non-Hermitian Hamiltonians \cite{12'}-\cite{18"}.

The adiabatic geometrical phase theory in a continuous spectrum was raised
for the first time by Maamache and Saadi \cite{11}-\cite{11'}. where they
extend the adiabatic geometrical phase theory for to the continuous spectra
and look at the eigenfunctions in a continuous spectrum $\left\vert \phi
_{\lambda }(t)\right\rangle $\ of the Hamitonian operator $H(t)$\ ($%
H(t)\left\vert \phi _{\lambda }(t)\right\rangle =E_{\lambda }(t)\left\vert
\phi _{\lambda }(t)\right\rangle $) and the solution $\left\vert \Psi
_{\lambda }(t)\right\rangle $\ of theSchr\"{o}dinger equation in the form 
\begin{equation}
\left\vert \Psi _{\lambda }(q;t)\right\rangle =e^{i\mu _{\lambda
}(t)}\left\vert \phi _{\lambda }\left( q;t\right) \right\rangle ,\text{ \ \
\ \ }
\end{equation}%
where $\langle \phi _{\lambda ^{\prime }}\mid \phi _{\lambda }\rangle
=\delta \left( \lambda ^{\prime }-\lambda \right) .$\ To find the phase, we
will substitute this solution into the Schr\"{o}dinger equation and then
project this equation onto the $\langle \phi _{\lambda }\left(
q_{l};t\right) \mid $\ (or the $\langle \phi _{\lambda ^{\prime }}\left(
q_{l};t\right) \mid $), one find that the phase is proportional to $\delta
\left( \lambda ^{\prime }-\lambda \right) $\ which, once integrated over $%
\lambda ^{\prime }$\ (or $\lambda $\ ) lead to the following practical
result 
\begin{equation}
\frac{d}{dt}\mu _{\lambda }(t)=\underset{-\infty }{\overset{+\infty }{\int }}%
\left\langle \phi _{\lambda ^{\prime }}\left( q_{l};t\right) \right\vert i%
\frac{\partial }{\partial t^{\prime }}-H\left( t^{\prime }\right) \left\vert
\phi _{\lambda }\left( q_{l};t\right) \right\rangle d\lambda ^{\prime }.
\end{equation}%
In everything that has been discussed so far, the eigenvalues are assumed to
be positive and discrete. A natural question arises: what would happen if
the eigenvalues were imaginary?

\section{Solution of the Schr\"{o}dinger equation}

We consider a quantum system described by a Hamiltonian $H(\overrightarrow{R}%
(t))$ that depends on a real time dependent parameter $\overrightarrow{R}(t)$
which parametrizes the environment of the system and such that $%
\overrightarrow{R}(T)=\overrightarrow{R}(0)$ . The time evolution is
described by the time dependent Schr\"{o}dinger equation (\ref{Sch}). At any
instant, the natural basis consists of the eigenstates $\left\vert n%
\overrightarrow{(R}(t))\right\rangle $ corresponding to the imaginary
energies $iE_{n}(t)$, i.e. such that%
\begin{equation}
H(\overrightarrow{R}(t))\left\vert n\overrightarrow{(R}(t))\right\rangle
=iE_{n}\overrightarrow{(R}(t))\left\vert n\overrightarrow{(R}%
(t))\right\rangle ,  \label{vp}
\end{equation}%
is fulfilled. \ 

The eigenfunctions of the Hamiltonian $H(\overrightarrow{R}(t))$ do not
generally belong to L$^{2}$\ and, as a result, do not form a complete set of
orthonormal functions $\langle m\overrightarrow{(R}(t)\mid n\overrightarrow{%
(R}(t)\rangle \neq \delta _{mn}$. A key question arising from the current
interest in Hermitian quantum mechanics with imaginary eigenvalues is: What
are the necessary and sufficient conditions for obtaining normalizable
solutions? To address this, we propose introducing a metric $\eta $\ ,
inspired by the pseudo-Hermitian framework, to define a scalar product that
guarantees a well-defined and positive norm,

\begin{equation}
\left\langle m\overrightarrow{(R}(t))\right\vert \eta (t)\left\vert n%
\overrightarrow{(R}(t))\right\rangle =\delta _{mn}.  \label{ps}
\end{equation}

Adiabatically, a system prepared in one of these states $\left\vert n%
\overrightarrow{(R}(0))\right\rangle $ will evolve with $H(\overrightarrow{R}%
(t))$ and so be in the state $\left\vert n\overrightarrow{(R}%
(t))\right\rangle $ at $t$.

Thus $\left\vert \Psi _{n}(t)\right\rangle $ can be written as 
\begin{equation}
\left\vert \Psi _{n}(t)\right\rangle =e^{\int_{0}^{t}E_{n}(t)dti}e^{\gamma
_{n}(t)}\left\vert n\overrightarrow{(R}(t))\right\rangle  \label{sol0}
\end{equation}%
where the Schr\"{o}dinger equation implies the following equation for $%
\gamma _{n}(t)$: 
\begin{equation}
\overset{\cdot }{\gamma }_{n}(t)=\left\langle n\left( \overrightarrow{R}%
(t)\right) \left\vert \eta (t)\frac{\partial }{\partial t}\right\vert
n\left( t\right) \right\rangle  \label{gp}
\end{equation}

\bigskip The total amplitude shift splits into two parts: 
\begin{equation*}
\phi =\underset{\text{dynamical amplidute}}{\underbrace{e^{%
\int_{0}^{t}E_{n}(t)dti}}}+\underset{\text{Geometrical amplitude}}{%
\underbrace{e^{\gamma _{n}(t)}}.}
\end{equation*}%
\bigskip\ \ The geometric quantity $\gamma _{\ast }(C)$, which only depends
on the trajectory traced in Hilbert space, \ is analogous to the renowned
Berry phase, associated with the cyclic adiabatic evolution along $C$.
Meanwhile, the first exponential represents the dynamical amplitude, similar
to the familiar dynamical phase.

What can be said about the orthonormalization relation of the evolved
states? Is the relation Eq.(\ref{ps}) still valid for $\left\vert \Psi
_{n}(t)\right\rangle $? The answer, of course, is negative. To address this
question, let us introduce an associated metric defined as: $\widetilde{\eta 
}(t)=$ $\eta (t)e^{-2\int_{0}^{t}E_{n}(t)dt}e^{-2\gamma _{n}(t)}$ . This
definition allows us to generalize the relations stated in equation (\ref{ps}%
) as follows:

\begin{equation}
\left\langle \Psi _{m}(t)\right\vert \widetilde{\eta }(t)\left\vert \Psi
_{n}(t)\right\rangle =\delta _{mn}
\end{equation}

\section{\protect\bigskip Example}

Consider the quantum system defined by the following Hamiltonian: 
\begin{equation}
H(t)=\frac{1}{2}\left[ Z\left( t\right) p^{2}+Y\left( t\right) \left(
pq+qp\right) +X\left( t\right) q^{2}\right] .  \label{cha1.16}
\end{equation}%
usually called a generalized harmonic oscillator which depends on the set \
of external parameters $(X,Y,Z)$ $\in 
\mathbb{R}
^{3}$. For fixed values of the parameters, $H$\ corresponds to the
Hamiltonian of the harmonic oscillator with purely imaginary frequency
provided $XZ$ $\langle \ Y^{2}$ and we assume henceforth that this remains
the case as $X,Y,Z$ vary.

The eigenvalue equation takes the following form 
\begin{equation}
\left[ -\frac{Z}{2}\frac{\partial ^{2}}{\partial q^{2}}-iYq\frac{\partial }{%
\partial q}+\left( \frac{Xq^{2}-iY}{2}\right) \right] \psi _{n}=E_{n}\psi
_{n},  \label{cha1.28}
\end{equation}

\bigskip To determine the eigenfunctions $\psi _{n}$ and the eigenvalues, we
introduce the following unitary transformation

\begin{equation}
\psi _{n}^{\mathrm{r}}(q)=U\psi _{n}(q)=\exp \left[ \frac{iY}{2Z}q^{2}\right]
\psi _{n}(q)  \label{cha3.12}
\end{equation}%
such as

\begin{equation}
U^{\dagger }qU=q,\text{ \ \ \ \ \ }U^{\dagger }pU=p-\frac{Y}{Z}q\text{ ,\ }
\label{cha3.12'}
\end{equation}%
and let us note that the transformed Hamiltonian $H^{\mathrm{r}}$ 
\begin{equation}
H^{\mathrm{r}}=U^{\dagger }HU=\frac{1}{2}\left[ Zp^{2}+\frac{(XZ-Y^{2})}{Z}%
q^{2}\right]  \label{r1}
\end{equation}%
corresponds to the Hamiltonian of the inverted harmonic oscillator with
purely imaginary frequency $i\sqrt{(Y^{2}-XZ)}=i\omega $.

Therefore, this replacement transforms the eigenfunctions $\psi _{n}$ into
generalized eigenvectors $\psi _{n}^{\mathrm{r}}$ of the inverted harmonic
oscillator (\ref{r1})

\begin{eqnarray}
\psi _{n}^{\mathrm{r}}(q) &=&\left( \frac{i\omega m}{\pi }\right) ^{\frac{1}{%
4}}\exp \left[ -\frac{i}{2}\frac{Y}{Z}q^{2}\right] \chi _{n}\left[ \left( 
\frac{i\omega }{2}\right) ^{\frac{1}{2}}q\right]  \notag \\
&=&\exp \left[ -\frac{i}{2}\frac{Y}{Z}q^{2}\right] V^{-1}\left\{ \left( 
\frac{\omega m}{\pi }\right) ^{\frac{1}{4}}\chi _{n}\left[ \left( \frac{%
\omega }{2}\right) ^{\frac{1}{2}}q\right] \right\}  \label{ch3.39}
\end{eqnarray}

where the operator $V=$ $\exp \left[ -\frac{\pi }{8}(qp+pq)\right] $ changes
the coordinate $x$ and momentum $p$ operators as%
\begin{align}
\exp \left[ \frac{\pi }{8}(qp+pq)\right] x\exp \left[ -\frac{\pi }{8}(qp+pq)%
\right] & =qe^{-i\frac{\pi }{4}},  \notag \\
\exp \left[ \frac{\pi }{8}(qp+pq)\right] p\exp \left[ -\frac{\pi }{8}(qp+pq)%
\right] & =pe^{i\frac{\pi }{4}},  \label{m}
\end{align}%
and its action on a wave function in the $q$ representation reads $Vf(q)=e^{i%
\frac{\pi }{8}}f(qe^{i\frac{\pi }{4}})$. \ The so-called $n$th Hermite
function $\chi _{n}\left( q\right) $ satisfies the following equation: 
\begin{equation}
\frac{\partial ^{2}\chi _{n}}{\partial x^{2}}+2\left( n+1-x^{2}\right) \chi
_{n}(x)=0,
\end{equation}%
and the corresponding energies are exactly given by

\begin{equation*}
E_{n}=(n+\frac{1}{2})\omega =i(n+\frac{1}{2})\sqrt{(Y^{2}-XZ)}
\end{equation*}

It is straightforward to verify that the norm associated with the solution (%
\ref{ch3.39}) is not well-defined. Specifically, when calculating the
squared norm of these functions, we observe that $\left\langle \psi _{n}^{%
\mathrm{r}}\right. \left\vert \psi _{n}^{\mathrm{r}}\right\rangle \neq 1$.
This clearly indicates that $\psi _{n}^{\mathrm{r}}$ does not belong to $%
L^{2}$.

This suggests to introduces the metric operator 
\begin{equation}
\eta =\exp \left[ -\frac{i}{2\hbar }\frac{Y}{Z}q^{2}\right] \exp \left[ 
\frac{\pi }{4}(xp+px)\right] \exp \left[ \frac{i}{2\hbar }\frac{Y}{Z}q^{2}%
\right] ,  \label{met}
\end{equation}%
such that the squared norm $\left\langle \psi _{n}^{\mathrm{r}}\right\vert
\eta \left\vert \psi _{n}^{\mathrm{r}}\right\rangle =1$ is finite. Note that
the metric operator is not unique, in the sense that infinitely many
operators can possibly implement.

Inserting the wave function (\ref{cha3.12}) and the metric (\ref{met}) into
the formula (\ref{gp}) defining the geometric quantity, \ one finds%
\begin{equation}
\gamma _{\ast }(C)=-(n+\frac{1}{2})\oint_{C}\frac{Z}{\omega }d\left( \frac{Y%
}{Z}\right)  \label{gm}
\end{equation}%
where the following property of the Hermite functions $\int_{-\infty
}^{+\infty }\zeta ^{2}\chi _{n}^{2}\left[ \zeta \right] d\zeta =(n+\frac{1}{2%
}),$has been used.

\section{Conclusion}

The adiabatic theorem\ concerns states\ $\left\vert \Psi
_{n}(t)\right\rangle $\ satisfying the time-dependent Schr\"{o}dinger
equation (\ref{Sch}) and asserts that if a quantum system with a
time-dependent non degenerate Hamiltonian $H\left( t\right) $\ is initially
in the $n$th eigenstates of $H\left( 0\right) $, and if $\ H\left( t\right) $%
\ evolves slowly enough, then the state at time $t$\ remains in the $n$th\
instantaneous eigenstates of $H\left( t\right) $\ up to a multiplicative
phase factor (\ref{sol0}) $\varphi _{n}\left( t\right) =\left\langle
n(t)\left\vert i\frac{\partial }{\partial t}-H(t)\right\vert
n(t)\right\rangle $, this, of course, is a completely standard, quantum
mechanical procedure. At first we have reviewed the basic elements of the
Berry phase and we have examined in detail the concept of geometrical
amplitudes in the context of a Hermitian Hamiltonian that admits imaginary
eigenvalues. In this paper, we introduced an approach to the so called
geometric amplitude. The paper is dedicated the original derivation of
geometrical amplitude , which points out that a system which evolves
cyclically under an adiabatic condition picks up an additional amplitude
factor which turns out to be geometrical in nature. The geometrical
amplitude concept is then shown on an example namely: a generalized harmonic
oscillator.

\bigskip I dedicates this work to my grandchildren Mohamed Chebouti ,
Taim-Abderahmane Maamache and Sadjed-Yazen Bendib


\begin{thebibliography}{99}
\bibitem{1} P. A. M. Dirac, The Principles of Quantum Mechanics, London:
Oxford University Press (1958).

\bibitem{Ben98} C. M. Bender and S.Boettcher, Real spectra in non-hermitian
hamiltonians having PT symmetry. Phys. Rev. Lett. 80, 5243 (1998).

\bibitem{Ben02} C. M. Bender, D. C.Brody and H. F Jones, Complex extension
of quantum mechanics. Phys. Rev. Lett. 89, 270401 (2002).

\bibitem{Mos02} A. Mostafazadeh, A. Pseudo-Hermiticity for a class of
nondiagonalizable Hamiltonians. J. Math. Phys. 43, 6343 (2002).

\bibitem{31} P. Ehrenfest, Bemerkung betreffs der spezifischen W%
\"{}%
arme zweiatomiger Gase,Verh. D. physik. Ges. 15, 451 (1913).

\bibitem{2} M. Born and v. Fock, Beweis des Adiabatensatzes\textrm{, }Z.
Phys. \textbf{51}, 165 (1928).

\bibitem{3} T. Kato, \ On the adiabatic theorem of quantum mechanics, J.
Phys. Soc.Jpn. 5, 435 (1950\textrm{)}.

\bibitem{4} A. Messiah, Quantum Mechanics (North-Holland, Amsterdam, 1962).

\bibitem{5} A. Galindo, P. Pascual, Quantum Mechanics (Springer 1991).

\bibitem{6} M.V. Berry, Quantal phase factors accompanying adiabatic
changes, Proc. R. Soc. A 392, 45 (1984).

\bibitem{8} Y. Aharonov, and J. Anandan, Phys. Rev. Lett., 58, 1593 (1987).

\bibitem{9} F. Wilczek and A. Zee, Phys. Rev. Lett. 52, 2111 (1984).

\bibitem{9'''} I. Bialynicki-Birula and Z. Bialynicka-Birula, Berry's phase
in the relativistic theory of spinning particles, Phys. Rev. D 35, 2383
(1987).

\bibitem{2"} M. Berry, The geometric phase, Scientific American 259, 46
(1988){\normalsize .}

\bibitem{10} J. Samuel, and R. Bhandari, Phys. Rev. Lett., 60, 2339 (1988).

\bibitem{3"} A. Shapere and F. Wilczek, \textquotedblleft Geometric Phases
in Physics\textquotedblright , World Scientific, Singapore (1989).

\bibitem{4"} H. Kuratsuji, Geometric Canonical Phase Factors and Path
Integrals, Phys. Rev. Lett. 61, 1687 (1988).

\bibitem{5"} G. Giavarini, E. Gozzi, D. Rohrlich and W. D. Thacker, On the
removability of Berry's phase, Journal of Physics A: Mathematical and
General 22, 3513 (1989).

\bibitem{4'''} M Maamache, J -P Provost and G Vallee, Berry's phase,
Hannay's angle and coherent states, J. Phys. A: Math. Gen. 23 5765 (1990).

\bibitem{6'''} J. W. Zwanziger, M. Koenig and A. Pines, Berry's Phase,
Annual Review of Physical Chemistry 41, 601 (1990).

\bibitem{5"'} A. Bohm, \textquotedblleft Quantum mechanics: Foundations and
Applications\textquotedblright , Springer, New York (1993)..

\bibitem{7"} J. Anandan, The geometric phase, Nature 360, 307 (1992).

\bibitem{7'''} M Maamache, J -P Provost and G Vallee, Berry's phase and
Hannay's angle from quantum canonical transformations, J. Phys. A: Math.
Gen. 24, 685 (1991).

\bibitem{8"} J. Anandan, The geometric phase, Nature 360, 307 (1992).

\bibitem{5"'} A. Bohm, \textquotedblleft Quantum mechanics: Foundations and
Applications\textquotedblright , Springer, New York (1993).

\bibitem{6a} J. Anandan, J. Christian and K. Wanelik, Resource Letter GPP-1:
Geometric Phases in Physics, Am. J. Phys. 65, 180 (1997).

\bibitem{14"} A. Tomita and R. Y. Chiao, Observation of Berry's Topological
Phase by Use of an Optical Fiber, Phys. Rev. Lett. 57, 937 (1986).

\bibitem{15"} R. Y. Chiao and Y. S. Wu, Manifestations of Berry's
Topological Phase for the Photon, Phys. Rev. Lett. 57, 933 (1986).

\bibitem{16"} T. Bitter and D. Dubbers, Manifestation of Berry's topological
phase in neutron spin rotation, Phys. Rev. Lett. 59, 251 (1987).

\bibitem{17"} Y. Hasegawa, R. Loidl, M. Baron, G. Badurek and H. Rauch,
Off-Diagonal Geometric Phase in a Neutron Interferometer Experiment, Phys.
Rev. Lett. 87, 070401 (2001).

\bibitem{18"} S. Filipp, Y. Hasegawa, R. Loidl and H. Rauch, Noncyclic
geometric phase due to spatial evolution in a neutron interferometer, Phys.
Rev. A 72, 021602 (2005).

\bibitem{19"} S. Filipp, J. Klepp et al., Experimental Demonstration of the
Stability of Berry's Phase for a Spin-1/2 Particle, Phys. Rev. Lett. 102,
030404 (2009).

\bibitem{20"} S. Sponar, J. Klepp et al., New aspects of geometric phases in
experiments with po- larized neutrons, Journal of Physics A: Mathematical
and Theoretical 43, 354015 (2010).

\bibitem{21"} D. Suter, G. C. Chingas, R. Harris and A. Pines, Berry's phase
in magnetic resonance, Mol. Phys. 61, 1327 (1987).

\bibitem{22"} D. J. Richardson, A. I. Kilvington, K. Green and S. K.
Lamoreaux, Demonstration of Berry's Phase Using Stored Ultracold Neutrons,
Phys. Rev. Lett. 61, 2030 (1988).

\bibitem{24"} Y. Hasegawa, M. Zawisky, H. Rauch and A. I. Ioffe, Geometric
phase in coupled neutron interference loops, Phys. Rev. A 53, 2486 (1996).

\bibitem{25"} C. L. Webb, R. M. Godun et al., Measurement of Berry's phase
using an atom interferometer, Phys. Rev. A 60, R1783 (1999).

\bibitem{26"} J. A. Jones, V. Vedral, A. Ekert and G. Castagnoli, Geometric
quantum computation using nuclear magnetic resonance, Nature 403, 869 (2000).

\bibitem{27"} D. Leibfried, B. DeMarco et al., Experimental demonstration of
a robust, high-fidelity geometric two ion-qubit phase gate, Nature 422, 412
(2003).

\bibitem{12'} J. C. Garrison and E. M. Wright, \ Complex geometrical phases
for dissipative systems, Phys. Lett. A 128, 177 (1988).

\bibitem{13'} G. Dattoli, R. Mignani and A. Torre, Geometrical phase in the
cyclic evolution of non-Hermitian systems,\ J. Phys. A: Math.Gen. 23, 5795
(1990).

\bibitem{14'} Ch. Miniatura, C. Sire, J. Baudon and J. Bellissard,
Geometrical phase factor for a non-Hermitian Hamiltonian, Europhys. Lett.
13, 199 (1990).

\bibitem{15'} A. Mondragon and E. Hernandez, Berry phase of a resonant
state, J. Phys. A 29, 2567 (1996).

\bibitem{16'} D. Viennot, A. Leclerc, G. Jolicard and J.P. Killingbeck,
Consistency between adiabatic and non-adiabatic geometric phases for
non-self-adjoint Hamiltonians, J. Phys. A: Math. Theor. 45, 335301 (2012 )..

\bibitem{18'} R. Hayward and F.Biancalana, Complex Berry phase dynamics in $%
\mathcal{PT}$-symmetric coupled waveguides, Phys. Rev. A. 98, 053833 (2018).

\bibitem{17} X.-C. Gao, J.-B. Xu, and T.-Z. Qian, Invariants and geometric
phase for systems with non-Hermitian time-dependent Hamiltonians, Phys. Rev.
A 46, 3626 (1992).

\bibitem{18} H. Choutri, M. Maamache, and S. Menouar, Geometric Phase for a
Periodic Non-Hermitian Hamiltonian, J. Korean Phys. Soc. 40, 358 (2002).

\bibitem{18"} S Cheniti, W Koussa, A Medjber and M Maamache, Adiabatic
theorem and generalized geometrical phase in the case of pseudo-Hermitian
systems, J. Phys. A: Math. Theor. 53, 405302 (2020).

\bibitem{11} M. Maamache, Y. Saadi, Adiabatic theorem and generalized
geometrical phase in the case of continuous spectra, Phys. Rev. Lett. 101,
150407 (2008).

\bibitem{11'} M. Maamache, Y. Saadi, \ Quantal phase factors accompanying
adiabatic changes in the case of continuous spectra, Rev. A 78, 052109
(2008).
\end{thebibliography}
\end{document}